\documentclass[twocolumn,showpacs,preprintnumbers,amsmath,amssymb]{revtex4}
\usepackage{graphicx}
\usepackage{dcolumn}
\usepackage{bm}

\pdfoutput=1

\newcommand{\lv}{\left \vert}
\newcommand{\rv}{\right \vert}
\newcommand{\la}{\left \langle}
\newcommand{\ra}{\right \rangle}
\newcommand{\ket}[1]{\lv #1 \ra}
\newcommand{\bra}[1]{\la #1 \rv}
\newcommand{\braket}[2]{\langle #1 \vert #2 \rangle}

\begin{document}

\title{Authorized quantum computation}

\author{Yu Tanaka}%
\affiliation{%
Department of Physics, Graduate School of Science, University of Tokyo, Tokyo 113-0033 Japan\\
Advanced Materials Laboratories, Sony, Kanagawa 243-0021 Japan
}%

\author{Mio Murao}
\affiliation{
Department of Physics, Graduate School of Science, University of Tokyo, Tokyo 113-0033 Japan\\
PRESTO, JST, Kawaguchi, Saitama 332-0012, Japan\\
Institute for Nano Quantum Information Electronics, University of Tokyo, Tokyo 153-8505, Japan
}%

\date{\today}

\begin{abstract}
We present authorized quantum computation, where only a user with a non-cloneable quantum authorization key can perform a unitary operation created by an authenticated programmer. The security of our authorized quantum computation is based on the quantum computational complexity problem of forging the keys from an obfuscated quantum gate sequence. Under the assumption of the existence of a \textit{sufficiently-random gate shuffling algorithm}, the problem is shown to be in the NQP (Non-deterministic Quantum Polynomial)-hard class by reducing it to a NQP-Complete problem, the exact non-identity check problem. Therefore, our authorized quantum computation can be computationally secure against attacks using quantum computers.
\end{abstract}

\maketitle


Consider the world once quantum computers exist and are widely used. In this world, unitary operations are the programs of quantum computers.  For the programmer of the quantum programs, they are important intellectual properties and it is important to protect the copyright of the programs.  On the other hand, for the user of the program, they do not need to know about the details of the program, they just want to perform a task, as long as the created by the authenticated programmer.

Such a situation can be solved if the programmer encodes a program so that the original program is only performable for users with non-cloneable authorization keys, distributes the encoded programs via an authenticator, and sends the authorization keys directly to the users.  Then, anyone can download the encoded programs via the authenticator, which are guaranteed to have been made by the programmer, but it is performable only by the authorized user.  In this letter, we propose a scheme of authorized quantum computation that allows this task in quantum computational security, as a possible new quantum cryptographic primitive.

In our scheme, we employ both quantum advantages and quantum limitations for its security. To protect the original programs from unauthorized users, the encoding process of the program is two-fold. One is an encryption process of introducing quantum authorization keys so that computation is not possible without using the correct key. The keys are unknown quantum states for any users (even for authorized users) and their anonymity and non-cloneability is ensured by quantum mechanics.  The other is an {\it obfuscation} process that hides the basis of the keys in the encoded program.  It is known that classically, obfuscating programs is impossible \cite{Barak}.  For quantum settings, existence of obfuscation with the help of quantum states is an open problem \cite{Aaronson}.  In this letter, by introducing the concept of gate shuffling algorithms, we show a sufficient condition for the obfuscation process where {\it forging} the quantum authorization keys is computationally difficult even using quantum computers.  We stress that we do not obfuscate the quantum program itself, but the identity of the quantum authorization keys.

We note that if we do not require the authenticity of the program, blind quantum computation \cite{BQC}, which aims to perform a unitary operation without revealing the identity of input states, can be used for similar tasks.  Blind quantum computation is a {\it two-party} protocol based on informational security and it requires multiple quantum/classical communications during computation.  In contrast, our authorized quantum computation is a {\it semi-public} protocol based on computational security to ensure the authenticity of the program and no communication is required during computation.

We first present a construction of authorized quantum computation to sketch our scheme.  We regard unitary operations to be quantum programs.  Note that we consider only polynomial quantum programs, namely, unitary operations represented by an array of a {\it polynomial} number $p(n)$ of elementary unitary gates. A unitary operation $U$ is described by a polynomial classical bit sequence $\{ 0,1 \}^*$.  We call this classical information as a quantum gate sequence of $U$ and denote it by $x(U)$. (Throughout this letter, we use capital letters to represent unitary operations and small letters to represent their quantum gate sequences.) Due to the non-uniqueness of the gate sequence representation for unitary operations, we consider a {\it set} of all quantum gate sequences for a unitary operation $U$ consisting of at most $p(n)$ elementary gate arrays and denote it by $g_{p(n)}(U)$, or simply $g(U)$ if the specification of a function $p(n)$ is irrelevant.  Since calculations of the matrix representation of a unitary operation require exponential computational power in terms of $n$, we have to rely on the polynomial gate sequence representation of unitary operations when $n$ is large.  Now consider a programmer who wants to encode a unitary operation $U_i$ (where $i$ is an index for specifying the unitary operation) acting on a $n$-qubit input Hilbert space $\mathcal{H}_{input}^{\otimes n}$.

\begin{figure}
\includegraphics[width=70mm,bb=0 0 623 255]{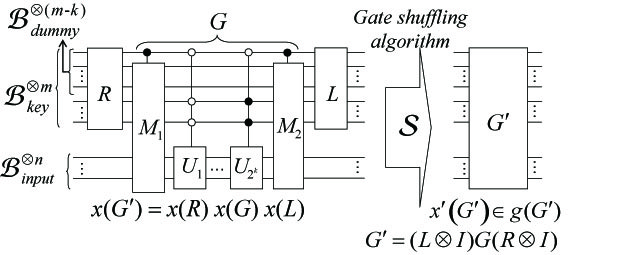}
\caption{A construction of gate sequences for the unitary
operations $G$ and $G^\prime$. See the text for notations.}
\label{fig:obfuscation}
\end{figure}

{\it Step 1:} The programmer extends the $U_i$ into another unitary operation $G$ acting on a larger Hilbert space $\mathcal{H}^{\otimes (m+n)}$ by adding a $m$-qubit Hilbert space $\mathcal{H}_{key}^{\otimes m}$ of quantum authorization keys in front of the input Hilbert space. (We often simply denote the quantum authorization key as the key.)  This extension is similar to the programmable quantum gate arrays proposed by Nielsen and Chuang \cite{PQC}.  The extended unitary operator $G$ transforms $G (\ket{i} \otimes \ket{\varphi}) = \ket{i} \otimes U_i \ket{\varphi}$, where $\{U_i \}$ is a set of $P=2^k$ $(1 \le i \le 2^k \ll 2^m)$ unitary operations for an arbitrary input state $\ket{\varphi} \in \mathcal{H}_{input}^{\otimes n}$, and $\{ \ket{i} \in \mathcal{H}_{key}^{\otimes m}\}$ is the corresponding key states in a computational basis specified by a binary bit sequence $\{ 0,1 \}^{k}$. The number of the key qubits $k$ should be taken to be of order $\log(n)$ for restricting the total gate number of $G$ to be in polynomial of $n$.  Thus the $(m-k)$-qubit {\it dummy} space $\mathcal{H}_{dummy}^{\otimes (m-k)}$ is introduced in the key space. A construction of $G$ for $\{ U_i \}$ is shown in Fig.~\ref{fig:obfuscation}, where $M_1$ and $M_2$ are random unitary operations acting on the dummy qubit space.

{\it Step 2:} By applying random unitary operations $L$ and $R$ on only $\mathcal{H}_{key}^{\otimes m}$ as $G \rightarrow G^\prime = (L \otimes I) G (R \otimes I)$ where $I$ denotes an identity operator of an appropriate dimension, we create a key state $\ket{\phi_i}=R^\dagger \ket{i}$ satisfying
\begin{eqnarray}
G^\prime (\ket{\phi_i} \otimes \ket{\varphi}) = \ket{\phi_i^\prime} \otimes U_i \ket{\varphi}, \label{sec}
\end{eqnarray}
where $\ket{\phi_i^\prime}=L \ket{\phi_i }$ is the key state after performing $G^\prime$. The programmer issues only one key for each authorized user. This step is the encryption process of the keys.

{\it Step 3:} The essence of our obfuscation process is in the non-uniqueness of the quantum gate sequence representation for unitary operations. The programmer transforms the series of quantum gate sequences $x(G^\prime)=x(R) x(G) x(L)$ into another obfuscated quantum gate sequence $x^\prime (G^\prime)$, where extracting information of $R$ and $L$ from $x^\prime (G^\prime)$ is not possible in a polynomial time even using quantum computers.  We call this transformation of quantum gate sequences as quantum gate shuffling. Later, we present a sufficient condition for a quantum gate shuffling algorithm for performing the obfuscation process.

{\it Step 4:} Classical information of the obfuscated gate sequence $x^\prime (G^\prime)$ is delivered to the trusted third party (authenticator) by the programmer.  It is authenticated and announced publicly.   On the other hand, the programmer directly sends the key $\ket{\phi_i}$ to an authorized user via a quantum channel. A pair of an authenticated public program $x^\prime$ and a set of authorization private keys $\{ \ket{\phi_i} \}$ is created.

{\it Step 5:} The user performs a unitary operation $G^\prime$ described by $x^\prime (G^\prime)$ on the joint state of the key $\ket{\phi_i}$ and an input state $\ket{\varphi}$ of user's choice.  Then the state $U_i \ket{\varphi}$ is obtained.  If the user does not use the correct key $\ket{\phi_i}$ and performs $G^\prime$, the resulting joint state cannot be a desired state and is highly likely to be entangled.  The security of our scheme will be discussed later.

We note that by performing the reverse quantum gate sequence using the used key state $\ket{\phi_i^\prime}$, the reverse unitary operation $U_i^\dagger$ can be also performed and the original key $\ket{\phi_i}$ is regained by ${G^\prime}^\dagger (\ket{\phi_i^\prime} \otimes \ket{\varphi})  = \ket{\phi_i} \otimes U_i^{\dagger} \ket{\varphi}$.  Therefore, we can recycle the quantum authorization key as long as it keeps coherence.

In this scheme, security depends on the obfuscation process given in {\it  Step 3}.  To present a sufficient condition for obfuscation, we investigate strategies for a malicious user, Eve.  We consider that Eve wants to perform the original unitary operation $U_i$ without using the quantum authorization key $\ket{\phi_i}$ issued by the programmer.  A powerful Eve may also be able to tap the quantum channels and steal other $N$ keys $\{ \ket{\phi_j} \}_{j=1,...,N}$ where $j \neq i$, and analyze the quantum gate sequence $x^\prime (G^\prime)$ to obtain a quantum gate sequence of $U_i$.  If the key is stolen, there is no way to prevent Eve from performing $U_i$, but we need to prevent Eve creating an unauthorized copy of the key.  We assume that Eve may destroy extra keys $\{ \ket{\phi_j} \}$ for $j \neq i$, but does not destroy the key $\ket{\phi_i}$ for performing $U_i$.  We regard that the key should be kept as the evidence of the authorized user.

Thus, we formally define authorized quantum computation implementing $2^k$ unitary operations $\{ U_i \}$ on $\mathcal{H}^{\otimes n}_{input}$ by the existence of a quantum gate sequence $x^\prime (G^\prime)$ and quantum authorization keys $\{ \ket{\phi_i} \}$ satisfying the following two conditions. {\bf{1}.}  $G^\prime$ is a unitary operation on an extended Hilbert space $\mathcal{H}^{\otimes (m+n)}$ satisfying Eq.~(\ref{sec}) for an arbitrary input state $\ket{\varphi} \in \mathcal{H}^{\otimes n}_{input}$.  {\bf{2}.} There is no polynomial quantum algorithm $\mathcal{A}$ such that
\begin{eqnarray}
\mathcal{A} : (x^\prime(G^\prime),\ \ket{\phi_i} \otimes \ket{\Phi_{i_N}})
\longmapsto
(y(F^\prime),\ \ket{\phi_i} \otimes \ket{\psi^y_{{x^\prime} i}}) \label{ext}
\end{eqnarray}
where $\ket{\Phi_{i_N}} = \ket{\phi_{j_1}} \otimes \cdots \otimes \ket{\phi_{j_N}}$ is a product state of $N$ keys ($i_N = j_1 \cdots j_N \in \{ 0,1 \}^{kN}$) and $y(F^\prime)$ is a quantum gate sequence of a unitary operation $F^\prime$, which allows $U_i$ to be performed by using a {\it forged} key state $\ket{\psi^y_{{x^\prime} i}} \in \mathcal{H}^{\otimes l}$ (for some integer $l$) as
\begin{eqnarray}
F^\prime (\ket{\psi^y_{{x^\prime}i}} \otimes \ket{\varphi}) = \ket{\psi^{\prime y}_{{x^\prime} i}} \otimes U_i \ket{\varphi}, \label{ext1}
\end{eqnarray}
for an arbitrary $\ket{\varphi} \in \mathcal{H}^{\otimes n}_{input}$ and $\bra{\psi^y_{{x^\prime} i}} \psi^y_{{x^\prime} j} \rangle=\delta_{ij}$.

Next, we investigate quantum gate shuffling algorithms for the obfuscation process.  Among algorithms mapping an element of $g_{p(n)}(U)$ to another element of $g_{q(n)}(U)$ where $p(n) \le q(n)$, we define a {\it completely-random shuffling} to be an algorithm randomly obtaining a quantum gate sequence from all possible quantum gate sequences of $g_{q(n)}(U)$.
To understand the power of random shuffling, we study {\it restricted} quantum gate sequences denoted by $z(C_I)$ of a controlled identity operation $C_{I}$ on $\mathcal{H}^{\otimes (n+1)}$ constructed by a quantum gate sequence $x(I)$.
For a given polynomial quantum gate sequence $x(U)$ of a general unitary operation $U$ on $\mathcal{H}^{\otimes n}$, we can always construct a corresponding quantum gate sequence $z(C_{U})$ of a controlled unitary operation $C_{U}$ on $\mathcal{H}^{\otimes (n+1)}$ by adding a control qubit in front of original qubits, replacing all the gate elements by controlled-gate operations and further decomposing them into elementary gate operations. This procedure can be completed in polynomial steps in $|x(U)|$.
Note that the restricted quantum gate sequence $z(C_I)$ can be also constructed from $x(e^{i\theta}I)$.\cite{phase}

We consider that a quantum gate sequence $x(I) \in g_{p(n)}(I)$ is given by a non-trivial combination of elementary gates and we further apply a quantum gate sequence of a unitary operation $V$ acting only on the control qubit (the first qubit) Hilbert space $\mathcal{H}_{control}$.  We compare the quantum gate sequences of $(V \otimes I) C_{I}$ and $C_{I}(V \otimes I)$.  If there exists a random shuffling algorithm in polynomial time, both sets are given by $g(V \otimes I)$ and they are identical.  Thus, after the random shuffling process, we cannot distinguish whether the quantum gate sequence of $V$ was originally applied from the right-hand side of the controlled identity or from the left-hand side. Information of the position of $V$ is lost.  This information loss is a key idea for our security proof.

However, the existence of a completely-random gate shuffling algorithm in polynomial time is not known and it is unlikely.  Instead, we introduce a concept of a sufficiently-random gate shuffling algorithm. To ensure informational indistinguishability of the applied order of $V$ on $\mathcal{H}_{control}$, shuffling algorithms are required that two sets of quantum gate sequences obtained by shuffling $z_l=x(V) z(C_I)$ and $z_r=z(C_I) x(V)$ should be overlapped significantly.  Thus we define a sufficiently-random gate shuffling algorithm as the following: A gate shuffling algorithm $\mathcal{S}$ in polynomial time is said to be sufficiently-random if for a restricted quantum gate sequence $z(C_I)$ constructed from arbitrary $x(I) \in g_{p(n)}(I)$ and a gate sequence $x(V)$ of $V$ on $\mathcal{H}_{control}$, there exists a polynomial function $q(n)$ for the number of qubit $n$ such that
\begin{eqnarray}
\frac{|\mathcal{D}(x(V) z(C_I)) \cap \mathcal{D}(z(C_I) x(V))|}{|\mathcal{D}(x(V) z(C_I)) \cup \mathcal{D}(z(C_I) x(V))|} = O(\frac{1}{q(n)}), \label{cond}
\end{eqnarray}
where $\mathcal{D}(\bullet)$ is a distribution obtained by applying $\mathcal{S}$ on a quantum gate sequence $\bullet$ in polynomial time.

Note that a sufficiently-random gate shuffling does not require a completely-random shuffling from the following reasons.
A set of $z(C_I)$ constructed from $x(I) \in g_{p(n)}(I)$ is a subset of $g_{p(n+1)}(I)$ and not uniform.
Thus, $\mathcal{D}(x(V) z(C_I))$ and $\mathcal{D}(z(C_I) x(V))$ are not necessarily required to be an uniform distribution of $g_{p(n+1)}(I)$. Further, $q(n)$ of Eq.(\ref{cond}) may depend on $z(C_I)$ and $x(V)$, since we require $q(n)$ to be just a polynomial function.


By assuming the existence of the sufficiently-random gate shuffling algorithm, we prove that it is quantum-computationally difficult for Eve to perform a cracking algorithm $\mathcal{A}$ defined by Eq.~(\ref{ext}).  We show that the quantum computational complexity of this task is in NQP (Non-deterministic Quantum Polynomial)-hard class by reducing it to a NQP-Complete problem, {\it the exact non-identity check problem} \cite{ENIC} of large unitary gate sequences.  The exact non-identity check problem is defined by the following. Let $x$ be a quantum gate sequence implementing a unitary operation $U$ with an ancilla system, decide whether $U$ is proportional to the identity operation, {\it i.e.}, $U = e^{i\theta}I$, or not. It is proven in Ref.~\cite{ENIC} that computational complexity of the exact non-identity check problem is NQP-Complete \cite{Kitaev}.  The class NQP is considered to be one of the natural extensions of the class NP to quantum computational complexity.

To apply the algorithm $\mathcal{A}$ to the exact non-identity check problem, we introduce a {\it modified} non-identity check problem by extending a quantum gate sequence $x(U)$ on $\mathcal{H}^{\otimes n}$ into a restricted quantum gate sequence $z(C_U)$ on $\mathcal{H}^{\otimes (n+1)}$. Then we apply two unitary operations $V_L$ and $V_R$ on $\mathcal{H}$ (the controlled qubit) from the left hand side and the right hand side of $C_U$.  The resulting operation is written by $C^\prime_U=(V_L\otimes I) C_U (V_R\otimes I)$.  Similarly to Eq.~(\ref{sec}), this operation transforms  $C^{\prime}_U (\ket{\phi_i} \otimes \ket{\varphi}) = \ket{\phi_i^\prime}\otimes U^{i}\ket{\varphi}$ for an arbitrary input state $\ket{\varphi} \in \mathcal{H}^{\otimes n}$, where $i \in \{ 0,1 \}$, $\ket{\phi_i}= {V_R}^{\dagger} \ket{i}$ and $\ket{\phi_i^\prime}= {V_L} \ket{i}$ for a {\it single}-qubit key state.  (Note that $U^i$ denotes the $i$th power $U$ and it is different from $U_i$.)  We state the modified non-identity check problem as the following: {\it Given a quantum gate sequence $x (U)$, decide whether the quantum gate sequence of $C^\prime_U$ is in $g(G_0^\prime) \equiv g(V_L V_R \otimes I)$ or $g(G_1^\prime) \equiv g((V_L \otimes I) C_U ({V_R} \otimes I))$ where $U \neq I$.}

We investigate the action of the algorithm $\mathcal{A}$ on a sufficiently-random shuffled gate sequence $z^\prime(C^\prime_U) \in \mathcal{D}( z(C^\prime_U))$. Since Eq.~(\ref{ext}) has to be also satisfied for $N$ key states $\ket{\Phi_{i_N}}={V_R}^{\dagger} \otimes ... \otimes  {V_R}^{\dagger} \ket{i_N}$ for $i_N = j_1 \cdots j_N \in \{ 0,1 \}^N$ and linearity of the algorithm, Eq.~(\ref{ext}) holds even if we replace the key state by any mixture of key states, namely, $\ket{\Phi_{i_N}}\bra{\Phi_{i_N}} \to \sum_{i_N} p_{i_N} \ket{\Phi_{i_N}}\bra{\Phi_{i_N}}$ where $\{ p_{i_N} \}$ is an arbitrary probability distribution.  For the case of a unitary operation $C^\prime_U$, the key space $\mathcal{H}_{key}$ does not contain the dummy space. Thus, we can replace the mixture by a completely mixed state $I/2^N$.  This means that Eve has no advantage from collecting extra keys, instead, she just needs to prepare $I/2^N$ by herself.  Thus, we can omit $\ket{\Phi_{i_N}}$ in Eq.~(\ref{ext}) without loss of generality, and the action of the algorithm $\mathcal{A}$ can be simplified to
\begin{eqnarray}
\mathcal{A}^{\prime} : (z^\prime (C^\prime_U),\ \ket{\phi_i}) \mapsto (y(F^\prime),\ \ket{\phi_i} \otimes \ket{\psi^y_{z^\prime i}}). \label{ext21}
\end{eqnarray}

Further, Eq.~(\ref{ext21}) can be represented by a CPTP map $\Lambda_{z^\prime} (\ket{\phi_i}\bra{\phi_i}) = \ket{\phi_i}\bra{\phi_i} \otimes \sum_y p^y_{{z^\prime} i} \ket{y} \bra{y} \otimes \ket{\psi^y_{{z^\prime}i}}\bra{\psi^y_{{z^\prime} i}}$, where $\sum_{y} p^y_{{x^\prime} i} = 1$ and $y \in \{ 0,1 \}^*$ is an abbreviation of the quantum gate sequence $y(F^\prime)$ for a unitary operation $F^\prime$ satisfying Eq.~(\ref{ext1}).  Using the Steinspring representation \cite{HayashiTextBook}, $\Lambda_{z^\prime}$ can be simulated by a unitary operation $W_{z^\prime}$ by adding an appropriate dimensional ancilla $\ket{\bar{0}}=\ket{0...0}$ as
\begin{eqnarray}
W_{z^\prime} (\ket{\phi_i} \otimes \ket{\bar{0}}) = \ket{\phi_i} \otimes \sum_{y} \sqrt{p^y_{{z^\prime} i}} \ket{y} \otimes \ket{\psi^y_{{z^\prime}i}} \otimes \ket{y {z^\prime} i}, \label{ext3}
\end{eqnarray}
where $\braket{y {z^\prime} i}{y^{\prime} {z^\prime} i} = \delta_{yy^{\prime}}$.

For $U \neq I$, note that
\begin{eqnarray}
\sum_{y,{y^\prime}} \sqrt{p^y_{{z^\prime} 0} p^{y^\prime}_{{z^\prime} 1}} \braket{y}{{y^\prime}} \braket{\psi^y_{{z^\prime}  0}}{\psi^{y^\prime}_{{z^\prime} 1}} \braket{y {z^\prime} 0}{{y^\prime} {z^\prime} 1}=0.
\end{eqnarray}
Applying $W_{z^\prime}$ on $\ket{\phi_+} \otimes \ket{\bar 0}$, where $\ket{\phi_+} = V_R^\dagger (\ket{0}+\ket{1})/{\sqrt{2}}$, and  tracing out the ancilla qubits, we obtain $\Gamma_{z^\prime} (\ket{\phi_+}\bra{\phi_+}) = {\rm tr}_a[W_{z^\prime} (\ket{\phi_+}\bra{\phi_+} \otimes \ket{\bar 0}\bra{ \bar 0}) W_{z^\prime}^{\dagger}] = I/2$. Thus,  for {\it all} ${V_L}$ and ${V_R}$ and all $z^\prime(C_U) \in g(G_1^{\prime})$, we have
\begin{eqnarray}
\bra{\phi_i}\Gamma_{z^\prime} (\ket{\phi_j}\bra{\phi_j})\ket{\phi_i} = \delta_{ij}, \label{con1} \\
\bra{\phi_+}\Gamma_{z^\prime} (\ket{\phi_+}\bra{\phi_+})\ket{\phi_+} = 1/2. \label{con2}
\end{eqnarray}

For $U = I$, note that $z(C^\prime_I) \in g(G_0^{\prime}) = g(V_L V_R \otimes I) = g({V_L}^{\prime} {V_R}^{\prime} \otimes I)$ for ${V_L} \neq {V_L}^{\prime}$ and ${V_R} \neq {V_R}^{\prime}$ where ${V_L} {V_R} = {V_L}^{\prime}{V_R}^{\prime}$. For $O(1/q(n+1))$ of the sufficiently-random shuffled quantum gate sequences of $\mathcal{S}( z(C^\prime_I))$, we cannot determine which $V_R$ is taken.  This property leads a contradiction if we assume that we cannot perform the exact non-identity check problem in polynomial time without using a witness state.

Under the impossibility of the exact non-identity check problem, the two probabilities given by Eqs.~(\ref{con1}) and (\ref{con2}) for $U=I$ should not be different more than $O(1/poly)$. By taking $V_R=V_L=H$, where $H$ denotes a Hadamard operation, we have the probabilities $\bra{+}\Gamma_{z^\prime} (\ket{+}\bra{+})\ket{+} = 1$ and $\bra{0}\Gamma_{z^\prime} (\ket{0}\bra{0})\ket{0} = 1/2$.  However, under the existence of sufficiently-random shuffling, we can also take $V^\prime_R =V^\prime_L=I$.  To satisfy the impossibility of the exact  non-identity check problem, the probabilities also have to satisfy $\bra{0}\Gamma_{z^\prime} (\ket{0}\bra{0})\ket{0} = 1$ and $\bra{+}\Gamma_{z^\prime} (\ket{+}\bra{+})\ket{+} = 1/2$, which leads contradiction to the previous results.

Thus, if there is an algorithm $\mathcal{A}$, we can decide whether an obfuscated quantum gate sequence $z^\prime (C_U)$, obtained by using sufficiently-random shuffling algorithm for a given quantum gate sequence $x(U)$, belongs to $g(G_0^{\prime})$ or $g(G_1^{\prime})$ by observing the difference in probabilities of above two cases by repeating the processes many (but polynomial) times. Then it is possible to check that the given quantum gate sequence $x(U)$ is an identity or not in polynomial time. However, the exact non-identity check problem has been shown to be NQP-complete and it is hard to solve in polynomial time without using the witness state. Therefore, Eve's cracking strategy $\mathcal{A}$, analyzing the quantum gate sequence $x^\prime (G^\prime)$ obfuscated by the sufficiently-random shuffling algorithm $\mathcal{S}$, is shown to be a computationally hard problem even using quantum computers.

In this letter, we propose authorized quantum computation, where only a user with a non-cloneable quantum authorization key can perform a unitary operation genuinely created by a programmer.  In our scheme, the unitary operation is encrypted into another unitary operation acting on a larger Hilbert space in the form of programmable quantum arrays proposed by Nielsen and Chuang. Further, the quantum gate sequence of the encrypted unitary operation is obfuscated by a sufficiently-random shuffling algorithm and then, it is authenticated and publicly announced. To perform the original unitary operation, the user needs to obtain an quantum authorization key, which is provided by the programmer to the authorized user, and then performs the obfuscated gate sequence together with the key.

The security of our authorized quantum computation is based on the quantum computational complexity of forging the quantum authorization key from the obfuscated quantum gate sequence.  Under the assumption of the existence of a sufficiently-random shuffling algorithm, we have shown that the problem is NQP-hard by reducing it to a NQP-Complete problem, the exact non-identity check problem of large quantum gate sequences. Therefore, our authorized quantum computation can be computationally secure against attacks using quantum computers.

{\it Acknowledgements:}  The authors thank M. R{\"o}tteler, M. Ukita, Y. Kawamoto and D. Markham for useful comments.  This work is partly supported by Special Coordination Funds for Promoting Science and Technology.



\begin{thebibliography}{99}

\bibitem{Barak} B. Barak, O. Goldreich, R. Impagliazzo, S. Rudich, A. Sahai, S. Vadhan, K. Yang, Advances in Cryptography -- CRYPTO'01, Lecture notes in Computer Science, Springer-Verlag, 2001.

\bibitem{Aaronson} S. Aaronson, {\it Ten Semi-Grand Challenges for Quantum Computing Theory}
http://www.scottaaronson.com/writings/qchallenge.html.

\bibitem{BQC} P. Arrighi and L. Salvail, Int. J. of Quantum Information {\bf 4}, 883 (2006); A. Broadbent, J. Fitzsimons and E. Kashefi, quant-ph/0807.4154.

\bibitem{PQC} M. A. Nielsen and I. L. Chuang, Phys. Rev. Lett. {\bf 79}, 321
(1997).


\bibitem{phase} Let $H$ and $P_i$ be a Hadamard operation and a phase shift with phase $2\pi/p_i$, where $p_i$ is a prime number and $p_i \neq p_j, (i\neq j)$. Considering two operations $P_n=\otimes_{i=1}^n P_i$ and $H_n=H^{\otimes n}$, one can show that $U=e^{i \theta}I$ iff $[H_n, U]=[P_n, U]=0$. Thus, define $U_1=UP_nU^{\dagger}P_n^{\dagger}$ and $U_2=UH_nU^{\dagger}H_n$ for a given unitary operation $U$, and replace $z(C_U)$ with $z(C_{U_1})$ and $z(C_{U_2})$.

\bibitem{ENIC} Y. Tanaka, quant-ph/0903.0675v1.

\bibitem{Kitaev} A. Y. Kitaev, A. H. Shen, and M. N. Vyalyi \textit{Classical and Quantum Computation}, Graduate Studies in Mathematics {\bf 47}, Am. Math.
Soc., Providence, Rhode Island, (2002).

\bibitem{HayashiTextBook} M. Hayashi, {\it Quantum Information}, Springer-Verlag, 2006.

\end{thebibliography}
\end{document}